\documentclass[10pt,twocolumn]{article}

\usepackage{graphicx}
\begin{document}

\title{Far UV excitation of hydrogen and light emission; applications in astrophysics.}

\author{Jacques Moret-Bailly, Physique, Universit\'e de Bourgogne, F-21000 Dijon France.}

\maketitle
\twocolumn[\begin{abstract}Assuming a spherical symmetry, the extreme UV emitted by a very hot source ionizes low pressure molecular hydrogen making a transparent bubble of H II (Protons and electrons). For an increase of radius, intensity of extreme UV and temperature decrease, so that the plasma contains more and more atoms. A spherical shell, mainly of neutral atoms (H I) appears.
If this shell is optically thick at Lyman frequencies of H I, it is superradiant and a competition of modes selects modes tangent to a sphere for which many atoms are excited. Thus, a shell of plasma emits, into a given direction, tangential rays showing a ring in which selected modes are brighter.
While at Lyman frequencies, absorption of rays emitted by the source excites the atoms able to amplify the superradiance, a more powerful amplification of superradiance results from an induced scattering of the radial beams, which extends to feet of lines and progressively to the whole spectrum. Thermodynamics says that the brightness of radial and tangential beams tends to be equal; if the solid angle of observation is much larger for the ring than for the source, almost the whole light emitted by the source is transferred to the rings, and the source becomes invisible. Paradoxically, a glow due to incoherent scattering and impurities around the source remains visible.
As the scattering decreases with the decrease of the radial intensity, the brightness of the ring decreases with radius.
These characteristics are found in supernova remnant 1987A.
\end{abstract}]

Pacs {42.65.Es, 42.50.Md, 95.30.Jx}

\noindent

\section{Introduction}
The aim of this paper is a theoretical, qualitative study of the interaction of light emitted by an extremely hot source, with a surrounding, low density, homogeneous cloud of initially cold hydrogen. As such systems may exist around stars, this model could be used in astrophysics.

A spherical symmetry is assumed around the centre of the source $O$. The density of the gas is supposed low, so that we need not take an index of refraction into account.

\medskip
Section \ref{nota} reminds well known optics to set the notations.

Section \ref{super} describes a superradiant emission of a spherical shell of excited neutral atomic hydrogen ($H_I$).

Section \ref{scat} describes a coherent scattering of bright light by this shell.

Section \ref{astro} shows similarities between this theoretical system and an observation of a supernova remnant.

\section{Notations.}\label{nota}
Set $a$ and $b$ two states of identical molecules of a gas, and $E_a < E_b$ the corresponding energies. Set $N_a$ and $N_b$ the populations of these molecules (number of molecules in an unit volume); at a temperature of equilibrium $T_{a,b}$, $N_a/N_b = \exp{(E_b-E_a)/kT_{a,b}}$. If the molecules are in a blackbody at temperature $T_n$, $T_{a,b}=T_n$. In the blackbody, Planck's law \cite{Planck,Nernst} correlates the amplitude of the electromagnetic field, the spectral brightness of a monochromatic beam with temperature $T_n$ and wavelength $\lambda = c/\nu = hc/(E_b-E_a)$. Out of a blackbody, the relation remains true provided that $T_n$ is replaced by $T_{a,b}$, defining the temperature of a monochromatic beam from its spectral brightness and its wavelength.

Suppose that two holes are drilled in the blackbody, small enough for a negligible perturbation of the blackbody, but defining, at a wavelength $\lambda$, a Clausius invariant at least equal to $\lambda^2$, so that the diffraction of a beam propagating through the holes may be neglected. The coherent amplification coefficient of a beam propagating through the holes is larger than 1 (true amplification) if its initial temperature is lower than $T_n$  , else lower than 1 (absorption).

If the output brightness of the beam does not depend on the path, this brightness corresponds to temperature  $T_n$, and the gas is said optically thick. 

Out of a blackbody, define a “column density” as the path integral of the density of a type of atoms, and a “spectral column density” as the path integral of $N_b - N_a$; following Einstein \cite{Einstein} the amplification of light is an exponential function of the spectral column density. In weak homogeneous sources, the total field remains close to the zero point field, so that the increase of field is nearly proportional to the column density, that is to the path. More precisely, assuming a constant amplification coefficient, the total field is an exponential function of the path, and the lines become sharper because their centres are more amplified than their feet; it is the superradiance. In a strong, optically thick source, the lines saturate, tending to temperature $T_{a,b}$ for any input temperature.  

\section{Superradiance in a spherical shell of atomic hydrogen.}\label{super}

Absorption of extreme UV (beyond Lyman region) emitted by the source ionizes the gas into a plasma of protons and electrons, transparent to light. We suppose that some energy is lost by radiation of lines by impurities or during collisions, so that the transparency is not perfect, the ionization is limited, ionized atoms make a nearly transparent spherical bubble. For an increase of radius R, intensity therefore absorption of extreme UV decrease, temperature decreases too, so that the proportion of neutral hydrogen atoms increases, becoming noticeable  around 50 000 K. Supposing that the decrease of temperature continues, around 10 000 K the atoms start to combine into molecules. Thus the density of atoms in an excited state is maximal on (at least) a sphere $\Sigma$.

 At Lyman lines frequencies, the interactions of light with atomic hydrogen are strong, the  oscillator strength being, for instance 0.8324 for $\alpha$ line. Study the spherical shell of excited atoms, supposing that it is optically very thick at the main Lyman frequencies, at least for beams crossing large spectral column densities in the shell. 

A spontaneous emission into a virtual beam of Clausius invariant $\lambda^2$, is, in the usual model, made of pulses which are elementary modes; these modes are coherently amplified up to temperature $T_{a, b}$; as photoexcitations are large and collisions rare, some  $T_{a, b}$ may be much larger than the temperature of the gas, some populations may be inverted; the amplification depletes the population of excited state $E_b$ (2P state for Ly$_\alpha$).

Supposing that the gas has a low density, the amplification coefficient is low; during a pulse, the space- and time-coherent increases of amplitude add along an assumed long path, while, without coherence it is the intensities of spontaneous emissions which would add; thus, the temperature of space-coherent, superradiant beams tends quickly to the temperature of the gas. The beams emitted nearly tangentially to $\Sigma$ depopulate the outer regions, forbidding a start of different emissions.

Evidently, the state of the atoms, therefore the radius of $\Sigma$ depend on all light-matter interactions.

\medskip
Into a given direction, the superradiant beams are inside a hollow cylinder whose base is a ring centred on the source. Observed modes are defined by the pupil of the observer (in astrophysics, the mirror of the telescope) and a just resolved region of the ring; an angular competition of modes making columns of light let appear some regions brighter.

\section{Stimulated scattering of radial beams by the superradiant beams.}\label{scat}
In the ring, neutral atoms de-excited by superradiance may be re-excited by absorption of Lyman lines radiated by the source, then de-excited again by superradiance; this process is weaker than a resonant stimulated scattering. 

\begin{figure}
\includegraphics[height=6 cm]{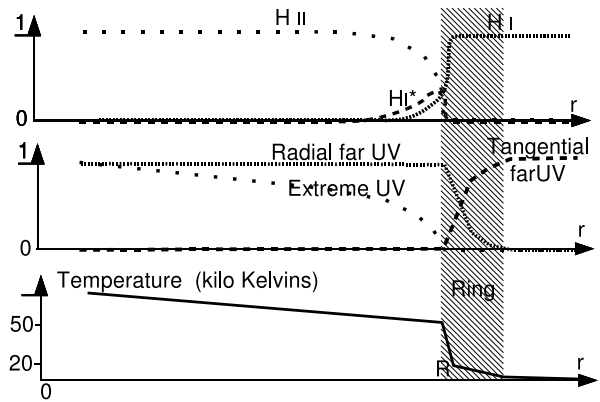}
\caption{\label{rad} Variation of the relative densities of H$_I$, H$_{II}$ and excited atomic hydrogen H$_I$*, relative intensities of light, and temperature along a radius of the shell of H$_I$.}
\end{figure}

A quasi resonant stimulated scattering of radial light emitted by the source may be induced by feet of the superradiant beams, starting  a permanent increase of linewidth of the stimulated lines. Finally, the whole spectrum may be transferred from the radial beams to the tangential beams. As the radial beams are progressively absorbed, the brightnesses of the tangential beams decrease from the inner rim of the observed ring.

The brightnesses of the tangential beams tend to be equal to the brightnesses of the radial beams; if the solid angle of observation of a ring is much larger than the solid angle of observation of the source, the intensity received by an observer from the ring tends to be much larger than the intensity received from the source which becomes invisible. On the contrary, if excited impurities and incoherent scattering emit a glow around the source, as the brightness of this glow is low, it is not absorbed, slightly amplified in the shell. 

The transfer of energy from radial to tangential beams starts at radius $R$ of $\Sigma$ where a fraction of hydrogen remains ionized. The depopulation of the excited states followed by collisions cools the gas, so that the proportion of neutral atoms increases more quickly than for  $r < R$ (figure \ref{rad}). This increase increases superradiance and scattering, increasing more the de-ionization...  . This reaction process may become catastrophic, transforming brutally all ions into neutral atoms, scattering a large fraction of the radial intensity, thus radiating very bright spots.

As the excited levels are strongly depopulated, it was sufficient to consider the strongest Lyman transitions. The emitted rays excite various mono- or poly-atomic molecules, generating long columns of excited molecules. These molecules radiate various superradiant lines in the direction of the columns, so that these lines seem having the same origin than the Lyman lines and the continuum; however, as the gas is colder than in the bubble, the lines are sharper.

\section{Application to astrophysics.}\label{astro}

The star of supernova remnant 1987A disappeared when an \textquotedblleft equatorial ring'' centered on it appeared \cite{apod} ; the density of neutral hydrogen increases mainly for a radius larger than 3/4 of the radius of the inner rim of the ring \cite{Lundqvist,Heng} and reaches a temperature of 50 000 K.

With densities of the order of $10^{10} m^{-3}$ and paths of the order of 0.01 light-year, that is column densities of the order of  $10^{24} m^{-2}$, this neutral atomic hydrogen is optically thick at Lyman frequencies \cite{Graves}.

The previous theoretical result may be a starting point for a mainly optical model of this remnant. Compare it, where applicable, to the standard model.

{ \it Main improvements:}

The standard model cannot explain how the star can disappear, while the glow which surrounds it remains visible \cite{Graves,Sonneborn,Lundqvist}. 

The brightness increases abruptly at the inner rim of the ring; in some places and time it starts so strongly that the name  \textquotedblleft hot spot'' \cite{apod} is used; then the decrease of brightness is more regular, so that the outer rim of the ring is not well defined. Angular variations of brightness, generating the  \textquotedblleft pearls of the necklace'' \cite{apod}, look like the modes observed in a conical, coherent emission of light.

{\it Remaining problems:}

The necklace is not circular, but elliptical, with irregularities: this may be explained by a larger density of the gas in direction of an axis z', and irregularities of the density \cite{Wang}.

There are two other, larger, outer rings. We may suppose that they result from the formation of two shells of H$_I$ around two neutron stars ejected along z' axis during a previous explosion of the star; an accretion of the gas could emit mainly extreme UV able to build shells, but light from the main star would be necessary to illuminate the shells.

\medskip
Measuring the times of propagation from the main star, along a path either direct, or with a scattering, the light-echoes method allows to locate the scattering matter. Thus, Sugerman et al. \cite{Sugerman} found two tumblers of matter which seem to be the shells generating the outer rings.

\section{Conclusion.}

 A theoretical model of interaction of light emitted by a very hot object with initially cold hydrogen, gives a repartition of light very similar to the observed images of supernova remnant 1987A. 

The standard model is more complex, its results are incomplete or less precise. Its computation of emission and scattering by incoherent interactions added in a Monte-Carlo process, close to Wolf's process \cite{Wolf}, does not work well \cite{Michael,Heng} because coherence must be taken into account.

Taking into account the stimulation of emissions, energized both by de-excitations and by scatterings, the origin of many rings observed in astrophysics could be explained from coherent optics and thermodynamics applied to simple models.

\end{document}